\documentclass[runningheads]{llncs}
\usepackage{graphicx}
\usepackage{times}

\usepackage{float}
\usepackage{epstopdf}
\usepackage{amsmath,amssymb,amsfonts,subfigure}
\usepackage{algorithm}
\usepackage{algorithmic}
\usepackage{cite}
\usepackage{amsmath,amsfonts,amssymb}

\usepackage{multicol}
\begin{document}
\title{Competitive Influence Propagation and Fake News Mitigation in the Presence of Strong User Bias}
\titlerunning{Competitive Influence Propagation and Fake News Mitigation}
%
\author{Akrati Saxena\inst{1} \and
Harsh Saxena\inst{2} \and
Ralucca Gera\inst{3}}
\authorrunning{Saxena et al.}
%
\institute{Department of Mathematics and Computer Science, \\
Eindhoven University of Technology, Netherlands, \email{a.saxena@tue.nl} \and
Department of CSE, KSVCEM Bijnor, India, \email{harshraj.saxena.18@gmail.com} \and
Department of Mathematics, Naval Postgraduate School, Monterey, CA, USA
\email{rgera@nps.edu}
}
\maketitle              
\begin{abstract}
Due to the extensive role of social networks in social media, it is easy for people to share the news, and it spreads faster than ever before. These platforms also have been exploited to share the rumor or fake information, which is a threat to society. One method to reduce the impact of fake information is making people aware of the correct information based on hard proof. In this work, first, we propose a propagation model called Competitive Independent Cascade Model with users' Bias (CICMB) that considers the presence of strong user bias towards different opinions, believes, or political parties. We further propose a method, called $k-TruthScore$, to identify an optimal set of truth campaigners from a given set of prospective truth campaigners to minimize the influence of rumor spreaders on the network. We compare $k-TruthScore$ with state of the art methods, and we measure their performances as the percentage of the saved nodes (nodes that would have believed in the fake news in the absence of the truth campaigners). We present these results on a few real-world networks, and the results show that $k-TruthScore$ method outperforms baseline methods. 

\end{abstract}
\section{Introduction}

\begin{figure*}
    \centering
    \includegraphics[width=12cm]{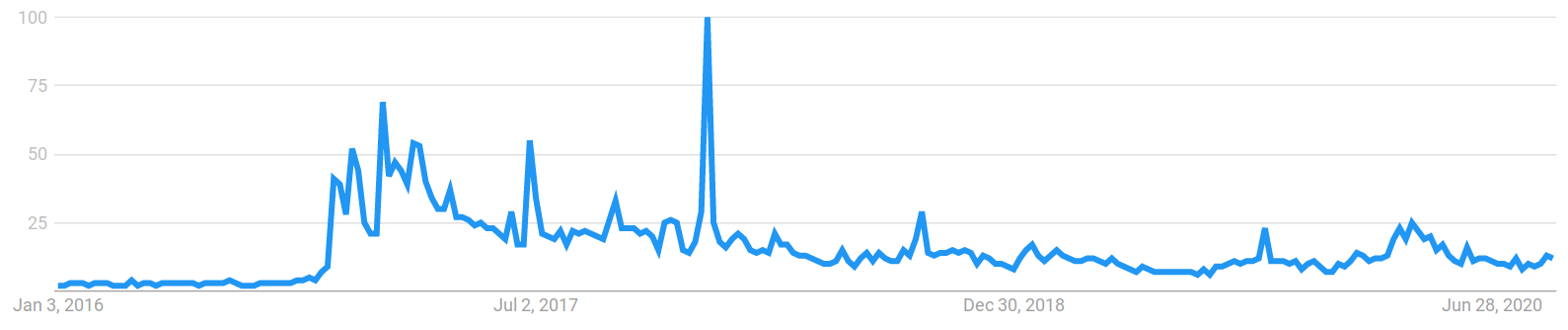}
    \caption{Google Trend for ``Fake News" web search since 2016.}
    \label{gtrend}
\end{figure*}

Since 1997, Online Social Networks (OSNs) have made it progressively easier for users to share the information with each other, and information reaches millions of people in just a few seconds. Over these years, people shared true as well as fake news or misinformation on OSNs, since no references or proofs are required while posting on an OSN. In 2017, The World Economic Forum announced that the fake news and misinformation is one of the top three threats to democracy worldwide~\cite{wef2017global}. Google Trend Analysis shows that the web search for the ``Fake News'' term began to gain relevance from the time of the U.S. presidential election in 2016~\cite{gt}; Figure \ref{gtrend} shows the plot we generated using Google Trend data. 

There are several reasons why people share fake news. Some of the threatening ones are changing the outcome of an event like an election, damaging the reputation of a person or company, creating panic or chaos among people, gaining profit by improving the public image of a product or company, etc. Less malicious reasons for sharing misinformation are due to the fame that users catch as a result of the news' catchiness or to start a new conversation while having no malicious intentions~\cite{chen2015social}. 

A study on the Twitter data shows that the false news spread faster, farther, and deeper \cite{vosoughi2018spread, park2012managing}, and these effects are even more prominent in the case of political news than financial, disaster, terrorism or science-related news \cite{vosoughi2018spread}. A large volume of fake information is shared by a small number of accounts, and Andrews et al. \cite{andrews2016keeping} show that this could be combated by propagating the correct information in the time of crisis; the accounts propagating true information are referred to as ``official" accounts.

In OSNs, users have high bias or polarity towards news topics, such as a bias for political parties \cite{soares2018influencers, wang2017polarized}. Lee et al. \cite{lee2018does} observe that the users who are actively involved in political discussions on OSNs tend to develop more extreme political attitudes over time than the people who do not use OSNs. Users tend to share the news confirming their beliefs. In this work, we propose a propagation model to model the spread of misinformation and its counter correct information in the presence of strong user bias; the proposed model is referred to as the Competitive Independent Cascade Model with users' Bias (CICMB). In the proposed model, the user's bias for a belief or opinion keeps getting stronger as they are exposed to more news confirming that opinion, and at the same time, their bias towards counter-opinion keeps getting weaken.

It is very challenging to mitigate the fake news in the presence of strong users' bias. Researchers have proposed various techniques to minimize the impact of fake news on a given social network. The proposed methods can be categorized as, (i) influence blocking (IB) techniques \cite{amoruso2017contrasting, pham2018targeted }, and (ii) truth-campaigning techniques (TC) \cite{budak2011limiting, song17}. IB techniques aim to identify a set of nodes that can be blocked or immunized to minimize the spread of fake information in the network. However, in truth campaigning techniques, the aim is to identify an optimal set of users who will start spreading the correct information in the network so that the people are aware of true news and share it further. Psychological studies have shown that people believe in true news rather than fake news when they receive both, and this also reduces the sharing of fake information further \cite{tanaka2013toward, ozturk2015combating}.

Most of the existing methods identify truth campaigners in the network who can minimize the impact of fake information; however, they do not consider the factor that a chosen node might not be interested in starting a truth campaign if asked \cite{budak2011limiting, nguyen2012containment, song17}. In this work, we consider a realistic approach where we have a given set of nodes which are willing to start a truth campaign; these nodes are referred to as \textit{prospective truth campaigners}. We propose a method to identify $k$ most influential truth campaigners from the given set of prospective truth campaigners to minimize the damage of fake news. We compare the proposed method, $k-TruthScore$, with state-of-the-art methods and the results show that the $k-TruthScore$ is effective in minimizing the impact of fake news in the presence of strong user bias.

The paper is structured as follows. In section~\ref{sec:RelatedWork} we discuss the related literature. In Section \ref{sec:cicmb}, we discuss the proposed spreading model. Section \ref{methodology} includes our methodology to choose truth-campaigners. Section \ref{results} shows the comparison of methods on real-world networks. We conclude the paper with future directions in Section \ref{conclusion}.

\section{Related Work}~\label{sec:RelatedWork}

The problem of fake news spreading needs public attention to control further spreading. In a news feed released by Facebook in April 2017~\cite{mosseri17}, Facebook outlined two main approaches for countering the spread of fake news: (i) a crowdsourcing approach leveraging on the community and third-party fact-checking organizations, and (ii) a machine learning approach to detect fraud and spam accounts. A study by Halimeh et al. supports the fact that Facebook's fake news combating techniques will have a positive impact on the information quality~\cite{halimehimpact}. Besides Facebook, there are several other crowdsourced fact-checking websites including \url{snopes.com}, \url{politifact.com}, and \url{factcheck.org}.

Researchers have proposed various influence blocking and truth-campaigning techniques to mitigate fake news in different contexts. In influence blocking, the complexity of the brute force method to identify a set of nodes of size $k$ to minimize the fake news spread is NP-hard \cite{amoruso2017contrasting}. Therefore, greedy or heuristic solutions are appreciated and feasible to apply in real-life applications. Amoruso et al. \cite{amoruso2017contrasting} proposed a two-step heuristic method that first identifies the set of most probable sources of the infection, and then places a few monitors in the network to block the spread of misinformation. 

Pham et al. \cite{pham2018targeted} worked on the Targeted Misinformation Blocking (TMB) problem, where the goal is to find the smallest set of nodes whose removal will reduce the misinformation influence at least by a given threshold $\gamma$. Authors showed that TMB is $\#P-hard$ problem under the linear threshold spreading model, and proposed a greedy algorithm that provides the solution set within the ratio of $1+ln(\gamma / \epsilon)$ of the optimal set and the expected influence reduction is greater than $(\gamma - \epsilon)$, given that the influence reduction function is submodular and monotone. Yang et al. worked on two versions of the influence minimization problem called Loss Minimization with Disruption (LMD) and Diffusion Minimization with Guaranteed Target (DMGT) using Integer Linear Programming (ILP) \cite{yang2019influence}. Authors proposed heuristic solutions for the LMD problem where $k$ nodes having the minimum degree or PageRank are chosen. They further proposed a greedy solution for the DMGT problem, where at each iteration, they choose a node that increases the maximal marginal gain.

In contrast to IB, truth campaigning techniques combat fake news by making the users aware of the true information. Budak et al. \cite{budak2011limiting} showed that selecting a minimal group of users to disseminate ``good" information in the network to minimize the influence of the ``bad" information is an NP-hard problem. They provided an approximation guarantee for a greedy solution for different variations of this problem by proving them submodular. Nguyen et al. \cite{nguyen2012containment} worked on a problem called $\beta^I_T$ where they target to select the smallest set $S$ of influential nodes which start spreading the good information, so that the expected decontamination ratio in the whole network is $\beta$ after $t$ time steps, given that the misinformation was started from a given set of nodes $I$. They proposed a greedy solution called Greedy Viral Stopper (GVS) that iteratively selects a node to be decontaminated so that the total number of decontaminated nodes will be maximum if the selected node starts spreading the true information. 

Farajtabar et al. \cite{farajtabar2017fake} proposed a point process based mitigation technique using the reinforcement learning framework. The proposed method was implemented in real-time on Twitter to mitigate a synthetically started fake news campaign. Song et al. \cite{song17} proposed a method to identify truth campaigners in temporal influence propagation where the rumor has no impact after its deadline; the method is explained in Section 5.2. In \cite{saxena2020mitigating}, authors considered users' bias, though the bias remains constant over time. In our work, we consider a realistic spreading model where users' biases keep getting stronger or weaken based on the content they are exposed to and share further. The competitive influence propagation is modeled by extending Independent Cascade Model (ICM). The ICM model has been extended to explain information propagation \cite{saxena2015understanding, kumari2017online, barbieri2013topic, gupta2019modeling}, and competitive influence propagation \cite{lin2019biog, song17, tong2017efficient, hosni2018least, simpson2018scalable, lv2017community} on OSNs; however users' bias are not considered while modeling. Next, we propose $k-TruthScore$ method to choose top-$k$ truth campaigners for minimizing the negative impact of fake news in the network.

\section{The Proposed Propagation Model: CICMB}\label{sec:cicmb}

The Independent Cascade Model (ICM) \cite{kempe2005influential} has been used to model the information propagation in social networks. In the existing ICM, each directed edge has an influence probability with which the source node influences the target node. The propagation is started from a source node or a group of source nodes. At each iteration, a newly influenced node tries to influence each of its neighbors with the given influence probability, and will not influence any of its neighbors in further iterations. Once there is no newly influenced node in an iteration, the propagation process is stopped. The total number of influenced nodes shows the influencing or spreading power of the seed nodes. 

Kim and Bock \cite{kim2011study} observed that peoples' beliefs construct their positive or negative emotions about a topic, which further affects their attitude and behavior towards the misinformation spreading. We believe that in real life, people have biases towards different opinions, and once they believe in one information, they are less willing to switch their opinion. 

\textbf{Competitive Independent Cascade Model with users' Bias (CICMB).} In our work, we propose a Competitive Independent Cascade Model with users' Bias (CICMB) by incorporating the previous observation in the ICM model when two competitive misinformation and its counter true information propagates in the network. In this model, each user has a bias towards misinformation at timestamp $i$, namely $B_m(u)[i]$, and its counter true information, $B_t(u)[i]$. The influence probability of an edge $(u,v)$ is denoted as, $P(u,v)$. Before the propagation starts, each node is in the neutral state, $N$. If the user believes in misinformation or true information, then it can change its state to $M$ or $T$, respectively. 

Once the propagation starts, all Misinformation starters will change their state to $M$, and truth campaigners will be in state $T$, and they will never change their state during the entire propagation being stubborn users. At each iteration, truth and misinformation campaigners will influence their neighbors as we explain next. A misinformation spreader $u$ will change the state of its neighbor $v$ at timestamp $i$ with the probability $Prob= P(u,v) \cdot B_m(v)[i]$. If the node $v$ change its state from $N$ to $M$, its bias values are updated as $(B_m(v)[i+1], B_t(v)[i+1])=f(B_m(v)[i], B_t(v)[i])$. Similarly, a truth campaigner $u$ will influence its neighbor $v$ with $P(u,v) \cdot B_t(v)[i]$ probability, and the bias values are updated as, $(B_m(v)[i+1], B_t(v)[i+1])=f(B_m(v)[i], B_t(v)[i])$. 

In our implementation, we consider that when a node $u$ believes in one information, its bias towards accepting other information is reduced in half, using these functions:\\
    (i). When $u$ changes its state to $M$, $B_m(u)[i+1]=B_m(u)[i]$ $\&$ $B_t(u)[i+1]={B_t(u)[i]}/2$.\\
    (ii). When $u$ changes its state to $T$, $B_m(u)[i+1]={B_m(u)[i]}/2$ $\&$ $B_t(u)[i+1]=B_t(u)[i]$.

\begin{figure}[t]
\centering
\includegraphics[width=5cm]{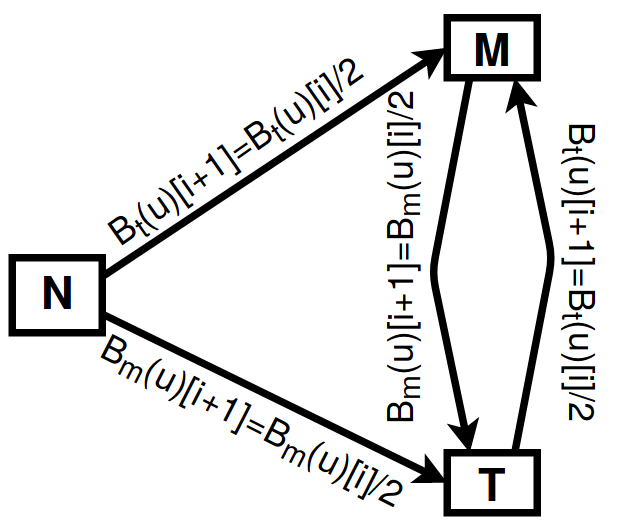}
\caption{CICMB Model used in the experiments, and the bias that is not displayed on the link stays constant between states.}
\label{icmb2}
\end{figure}

The model is explained in Figure \ref{icmb2} for this linear function. As a second case for our research, we will present results for another stronger bias function, where the bias is reduced faster, using a quadratic function. 

\section{Methodology}\label{methodology}

We first introduce the problem formulation and then follow with our proposed solution.

\subsection{Problem Formulation}

In OSNs, when true or misinformation is propagated, users change their state between the elements of the set $\{N, M, T\}$. The state of a user $u$ at timestamp $i$ is denoted as $\boldsymbol{\pi}_u[i]$ with the following possible assignments: (i) $\boldsymbol{\pi}_u[i]=T$ if user believes in true information, (ii) $\boldsymbol{\pi}_u[i]=M$ if user believes in misinformation, and (iii) $\boldsymbol{\pi}_u[i]=N$ if user is in neutral state. 

Given a set of rumor starters $R$ who spreads misinformation, the deadline of misinformation spread $\alpha$, and a set of prospective truth campaigners $P$, we aim to identify a set $D$ of chosen truth-campaigners of size $k$ from set $P$ ($D \subset P$ and $|D|=k$) to start a truth-campaign such that the impact of misinformation is minimized.  

Let $u$ be a node such that $\boldsymbol{\pi}_{u}[\alpha]= M$ if only misinformation is propagated in the network and $\boldsymbol{\pi}_{u}[\alpha] = T$, when both misinformation and its counter true information is propagated in the network. The node $u$ is considered a \emph{saved node} at the deadline $\alpha$ as it believes in true information and would have believed in the misinformation in the absence of truth-campaign.

\smallskip
\noindent \textbf{Problem Definition:}
Given a graph $G=(V, E)$, a rumor deadline $\alpha$, a set of rumor starters $R$, and a set of prospective truth-campaigners $P$. Let $S$ be the set of nodes whose state is $M$ at time $\alpha$ when only nodes in the set $R$ propagate misinformation using CICMB. Let $I$ be the set of nodes whose state is $T$ at time $\alpha$ when sets $R$ and $D$ propagate misinformation and true information, respectively, using CICMB. Our aim is to find a set $D \subset P$ of given size $k$, such that the number of saved nodes is maximized as follows:
\begin{eqnarray*}
f(D,M) &= & \sum_{v \in S \cap I} (1 | \boldsymbol{\pi}_{v}(\alpha) =T)
\end{eqnarray*}

\subsection{The Proposed Solution}

In this section, we introduce our proposed algorithm, $k-TruthScore$, giving intuition for how it works, and we then summarize it at the end of the section. For a given set of misinformation starters $R$ and prospective truth-campaigners $P$, our goal is to estimate which truth campaigner node will save the maximum number of nodes by the deadline $\alpha$ tracked by their TruthScore that we introduce below. We then choose top-$k$ nodes having the highest TruthScore as truth-campaigners ($D$) to minimize the impact of misinformation. 

To compute TruthScore, we assign to each node $u$, two arrays $mval$ and $tval$, each of length ($\alpha$ + 1), where $mval_u[i]$ and $tval_u[i]$ denote the estimated probability that node $u$ will change its state to $M$ and $T$ at time $i$, respectively. To estimate these probability values, first, we create the Directed Acyclic Graph (DAG) $G'(V, E')$ of the given network $G$ to remove the cycles from the network. Otherwise, if there would be a cycle in the network, then the nodes belonging to the cycle will keep updating the probabilities of each other in an infinite loop.

We now compute the probability of an arbitrary node $u$ changing its state to $M$ at some iteration $i$, namely $mval_u[i]$.  For this to happen, we compute two probabilities: 
\begin{enumerate}
    \item the probability that the node $u$ is not in state $M$ at time $i-1$ is computed as, $(1-\sum_{j=1}^{i-1}mval_u[j])$, and     
    \item the probability that the node $u$ will receive the misinformation at the $i^{th}$ step that considers all parents $v$ of node $u$ that have updated their $mval$ at $i-1$ timestamp, $V_1=\{v | (v,u) \in E' \; \& \; mval_v[i-1] > 0 \}$. Then we compute the value of $mval_u[i]$ by taking their product as shown in Equation~\ref{mval}:
    \begin{equation}\label{mval}
        mval_u[i]=\sum_{v \in V_1}(mval_u[i]+(1-mval_u[i])\cdot(1-\sum_{j=1}^{i-1}mval_u[j]) \cdot P(v,u)\cdot B_m(u) \cdot mval_v[i-1])
    \end{equation}
\end{enumerate}

We use this formula to compute $mval_u[i]$ for all nodes from $i=1$ to $\alpha$. All the nodes whose $mval$ has been updated, are added to set $A$.

Next, we compute the TruthScore of each prospective truth-campaigner $w$. We estimate the probability that a node $u$ will believe in true information at $i_{th}$ timestamp when the true information is propagated from node $w$ in the network. For this update $tval_w[0]=1$, and compute for each node $u \in (V-R)$, $tval_u[i]$ from $i=1$ to $\alpha$.

The probability that node $u$ will change its state to $T$ at timestamp $i$ is the probability that the node $u$ has not changed its state to $T$ at any previous timestamp multiplied by the probability of receiving the true information at $i_{th}$ timestamp. It is computed using the same approach as defined in Equation \ref{mval}.

The estimated probability that a node $u$ will change its state to $T$ at time stamp $i$ is computed as follows: Consider all parents $v$ of node $u$ who has updated $tval_v[i-1]$ at $i-1 $ timestamp, $V_2=\{v | (v,u) \in E' \; \& \; tval_v[i-1] > 0 \}$.
    \begin{equation}\label{tval}
        tval_u[i]=\sum_{v \in V_2}(tval_u[i]+(1-tval_u[i])\cdot(1-\sum_{j=1}^{i-1}tval_u[j]) \cdot P(v,u) \cdot B_t(u)\cdot tval_v[i-1])
    \end{equation}

The $tval$ is computed for $i=1$ to $\alpha$. All the nodes whose $tval$ has been updated, are added to $B$.
The truth score of truth-campaigner $w$ is computed as:
\begin{equation}\label{eqts}
TruthScore(w)=\sum_{v \in A \cap B}\sum_{i=1}^{\alpha}tval_v[i]
\end{equation}

For the fast computation, a node $v$ will update the $mval$ of its child node $u$ at timestamp $i$, if $mval_v[i-1] > \theta$, where $\theta$ is a small threshold value. The same threshold value is used while computing $tval$ array of the nodes. 

\vspace{.5cm}
We now summarize the above described method, and we call it  \textbf{$k$-TruthScore}:
\begin{enumerate}
    \item Create $G'(V, E')$, the DAG of the given network $G$.
    \item For all nodes in the set $R$ of rumor starters, update $mval_u[0]$=1. Compute $mval$ for the nodes reachable from $R$ by the given deadline $\alpha$ using Equation \ref{mval} and add these nodes to set $A$.
    \item For each given prospective truth-campaigner $w$ from set $P$,
    \begin{enumerate}
        \item Update $tval_w[0] =1$
        \item Compute $tval$ arrays for the nodes reachable by $w$ by the given deadline $\alpha$ using Equation \ref{tval}, and add these nodes to set $B$.
        \item Compute $TruthScore(w)$ by adding the values of $tval$ for the nodes in $A \cap B$ using Equation \ref{eqts}.
    \end{enumerate}
    \item Choose top-$k$ truth-campaigners having the highest $TruthScore$.
\end{enumerate}

The complete method is explained in Algorithm 1. 

\begin{algorithm}[!]
\caption{$k-TruthScore(G(V,E),R,P,k,\alpha,\theta)$}
\label{algo1}
\begin{algorithmic}
\STATE $G'(V,E')=$ Create DAG of $(G(V,E))$
\STATE Each node $u$ has $mval$ and $tval$ arrays of size $\alpha + 1$ initialized with $0$ entries.
\STATE $A=\{ \}$, and $B=\{ \}$.
\FOR{each node $u$ in $R$}
    \STATE $mval_u[0]=1$
\ENDFOR
\STATE $i=1$, $R' = R$
\FOR{$i$ in range($1$,$\alpha$ + 1)}
    \STATE $X=\{ \}$
    \WHILE{$R'$ is not empty}
        \STATE Remove node $v$ from $R'$
        \IF{$mval_v[i-1]> \theta$}
        \FOR{each child $u$ of $v$} 
        \STATE $mval_u[i]=mval_u[i]+(1-mval_u[i])\cdot(1-\sum_{j=1}^{i-1}mval_u[j])\cdot P(v,u)\cdot B_m(u)\cdot mval_v[i-1]$
        \STATE Add $u$ to $X$
        \STATE Add $u$ to $A$
        \ENDFOR
        \ENDIF
    \ENDWHILE
    \STATE $R'= X$
\ENDFOR 

\FOR{each node $w$ in $P$}
    \STATE $B$=\{ \}, $P'$=\{ \}
    \STATE $tval_w[i]=1$
    \STATE Add $w$ to $P'$
    \FOR{$i$ in range$(1,\alpha + 1)$}
        \STATE $X=$ \{ \}
        \WHILE{$P'$ is not empty}
            \STATE Remove node $v$ from $P'$
            \IF{$tval_v[i-1]> \theta$}
            \FOR{each child $u$ of $v$} 
                \IF{$u \notin R $}
                \STATE $tval_u[i]=tval_u[i]+(1-tval_u[i])\cdot(1-\sum_{j=1}^{i-1}tval_u[j])\cdot P(v,u)\cdot B_t(u)\cdot tval_v[i-1]$
                \STATE Add $u$ to $X$
                \STATE Add $u$ to $B$
                \ENDIF
            \ENDFOR
            \ENDIF
        \ENDWHILE
        \STATE $P'$=$X$
    \ENDFOR
    
    \FOR{each node $u$ in $A \cap B$}
        \FOR{$i$ in range $(1, \alpha + 1)$}
            \STATE $TruthScore(w)=TruthScore(w)+tval_u[i]$
        \ENDFOR
    \ENDFOR
    \STATE Update all entries of $tval$ array for all nodes with $0$.
\ENDFOR
\STATE $D$ = choose top-$k$ nodes from set $P$ having highest TruthScore
\STATE Return $D$
\end{algorithmic}
\end{algorithm}

\section{Performance Study}\label{results}

We carry out experiments to validate the performance of the $k-TruthScore$ to identify top-$k$ truth-campaigners. 

\subsection{Datasets}

We perform the experiments on three real-world directed social networks, Digg, Facebook, and Twitter, as presented in Table~\ref{dataset}. For each of them, the diameter is computed by taking the undirected version of the network.

\begin{table}[H]
\centering
\begin{tabular}{|l|l|l|c|c|}
\hline
Network	&	Nodes	&	Edges	& Diameter &	Ref	\\ \hline
Digg	&	29652	&	85983	& 12 &	\cite{de2009social}	\\ \hline
Facebook	&	43953	&	262631	& 18 &	\cite{viswanath2009evolution}	\\ \hline
Twitter &	81306	&	1768135	& 7 &	\cite{de2010does}	\\ \hline
\end{tabular}
\caption{Datasets}
\label{dataset}
\end{table}

We assign the influence probability of each edge $(v,u)$ uniformly at random (u.a.r.) from the interval $(0, 1]$. Each node in the network has two bias values, one for the misinformation and another for the true information. For misinformation-starters, the bias for misinformation is randomly assigned a real value between $[0.7, 1]$ as the nodes spreading misinformation will be highly biased towards it. For these nodes, the bias for true information will be assigned as, $B_t[0] = 1-B_m[0]$. 

Similarly, the nodes chosen to be prospective truth campaigners will have a high bias towards true information, and it will be assigned u.a.r. from the interval $[0.7, 1]$. For prospective truth-campaigners, the bias for misinformation will be assigned as, $B_m[0] = 1-B_t[0]$. For the rest of the nodes, the bias value for misinformation and their counter true-information will be assigned uniformly at random from the interval $(0, 1]$. Note that the size of the prospective truth-campaigners set is fixed as $|P|=50$, and set $P$ is chosen u.a.r. from set $(V-R)$. We fix $\theta=0.000001$ for all the experiments.

\subsection{Baseline Methods}

We have compared our method to the following two state-of-the-art methods. 

\begin{enumerate}
	\item \textbf{Temporal Influence Blocking (TIB) \cite{song17}.} The TIB method runs into two phases. In the first phase, it identifies the set of nodes that can be reached by misinformation spreaders by the given deadline. Then, it identifies the potential nodes that can influence these nodes. In the second phase, it generates Weighted Reverse Reachable (WRR) trees to compute the influential power of identified potential mitigators by estimating the number of reachable nodes for each potential mitigator. In our experiments, we select the top-$k$  nodes to be the prospective truth-campaigners having the highest influential power.
	\item \textbf{Targeted Misinformation Blocking (TMB) \cite{pham2019minimum}.}  The TMB computes the influential power of a given node by computing the number of saved nodes if the given node is immunized in the network. Therefore, the influence reduction of a node $v$ is computed as $h(v) = N(G) - N(G \setminus v)$, where $N(G)$ and $N(G \setminus v)$ denote the number of nodes influenced by misinformation starters in the $G$ and $(G \setminus v)$, respectively. We then select top-$k$ nodes having the highest influence reduction as truth-campaigners.   \\
	
	After selecting top-$k$ truth campaigners using TIB and TMB methods, the CICMB model is used to propagate misinformation and counter true information.
\end{enumerate}

If set $R$ starts propagating misinformation, then $S$ is the set of nodes whose state is $M$ at $t = \alpha$. If set $R$ propagates misinformation, and set $D$ propagates true information, then let $I$ be the set of nodes whose state is $T$ at $t = \alpha$. The performance of various methods is evaluated by computing the percentage of nodes saved, i.e., $\frac{|S|-|I|}{|S|}\cdot100$.

We compute the results by choosing five different sets of misinformation starters and truth-campaigners. In several instances, each experiment is repeated $100$ times, and we report their average value to show the percentage of saved nodes.

\begin{figure*}
\centering
\begin{minipage}[t]{.47\textwidth}
\includegraphics[width=\linewidth]{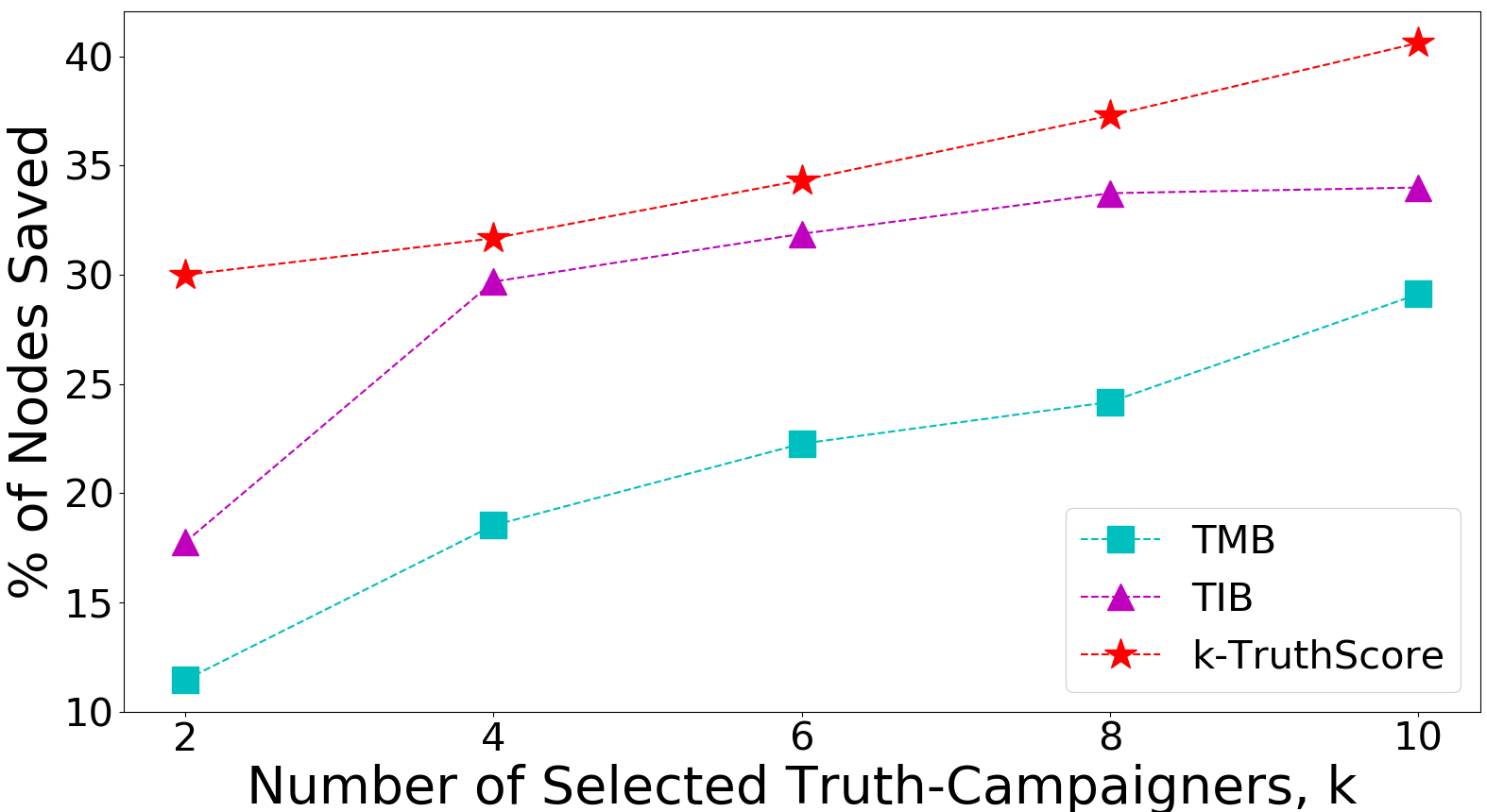}
	\begin{center} { (a) Digg}\end{center}
	\includegraphics[width=\linewidth]{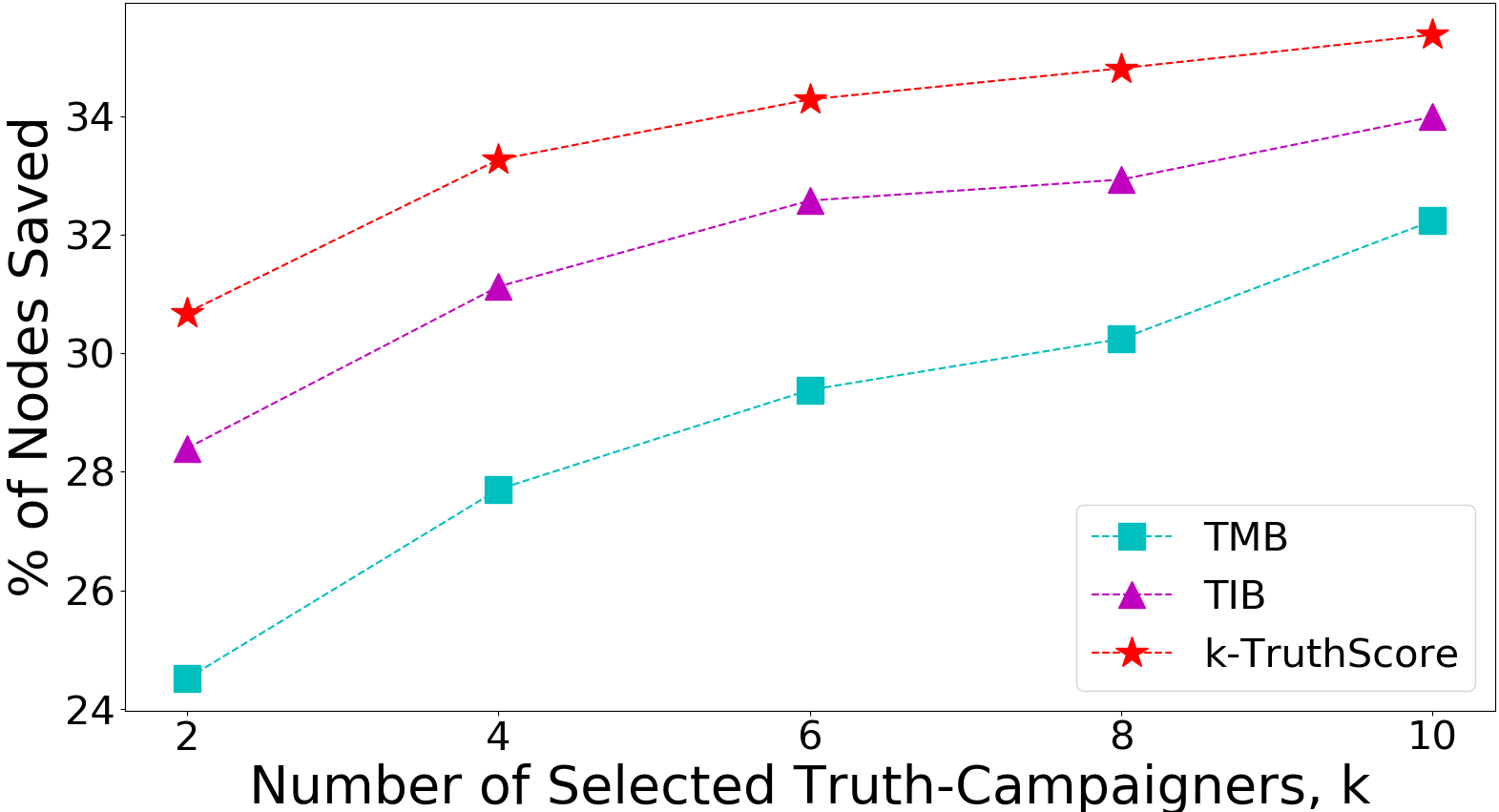}
	\begin{center} { (b) Facebook}\end{center}		
	\includegraphics[width=\linewidth]{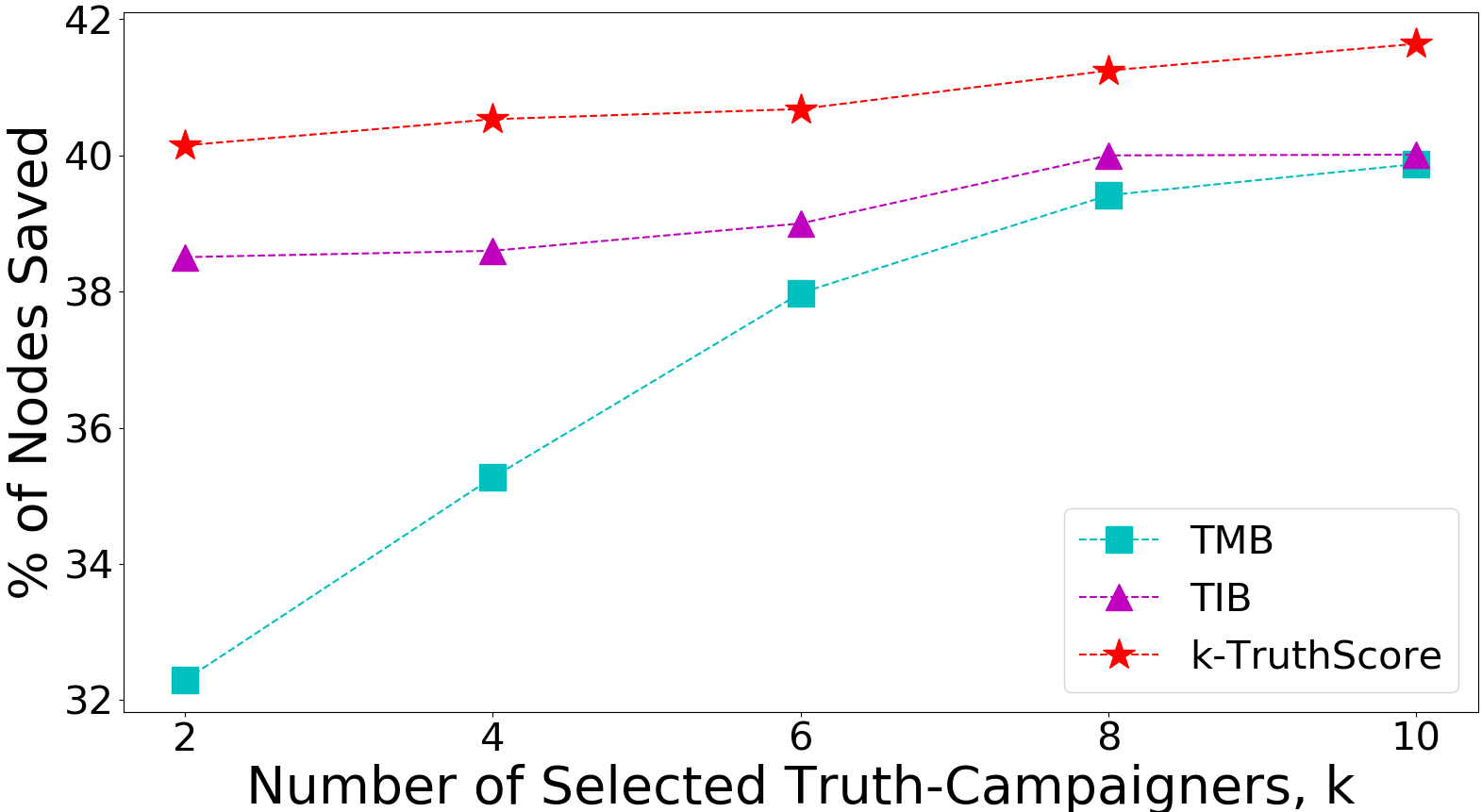}
	\begin{center} { (c) Twitter}\end{center}	
	\caption{Effect of varying $k$ for node selection methods.}
\label{fig1a}
\end{minipage}\qquad
\begin{minipage}[t]{.47\textwidth}
\includegraphics[width=\linewidth]{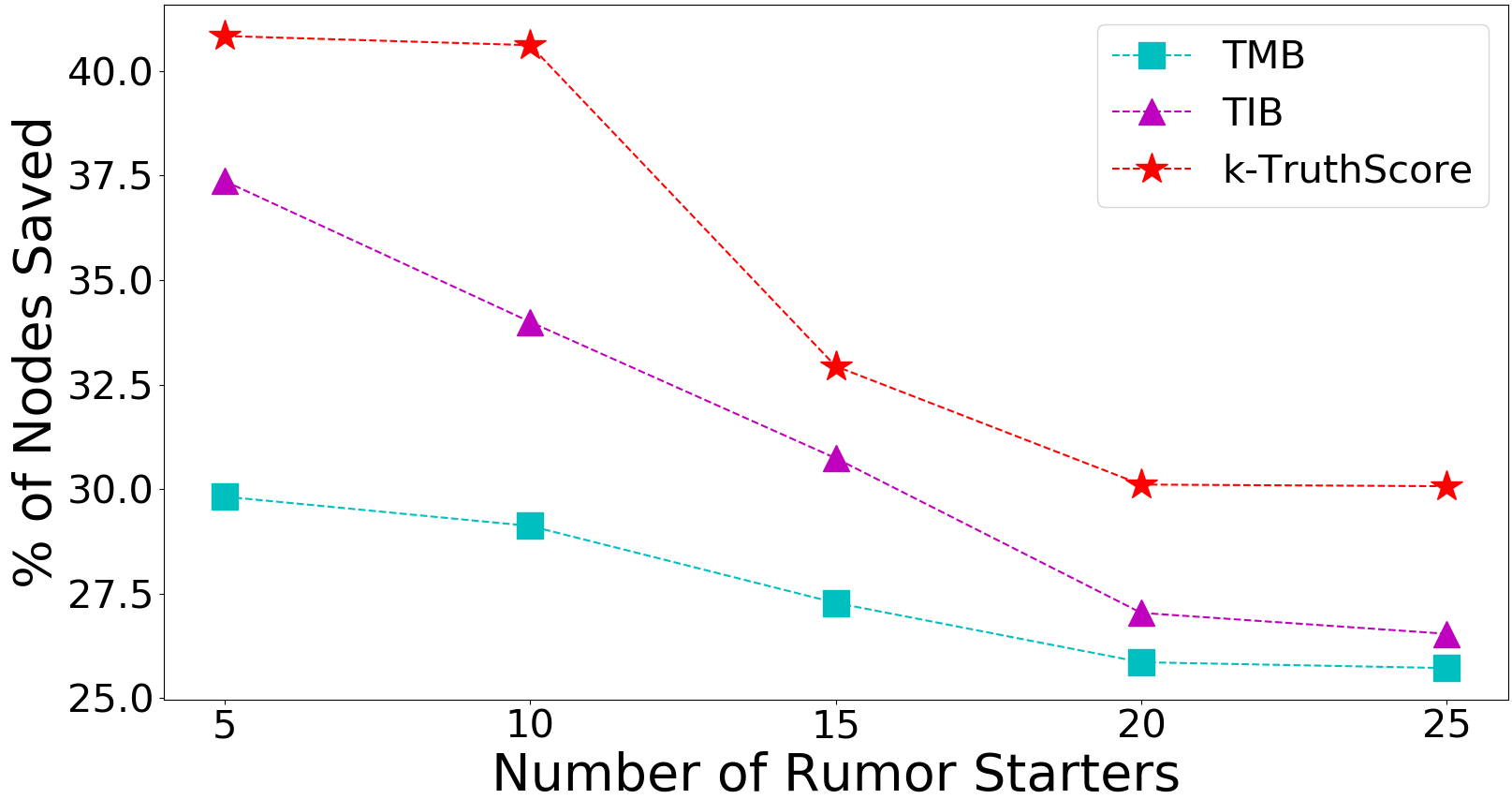}
	\begin{center} { (a) Digg}\end{center}
	\includegraphics[width=\linewidth]{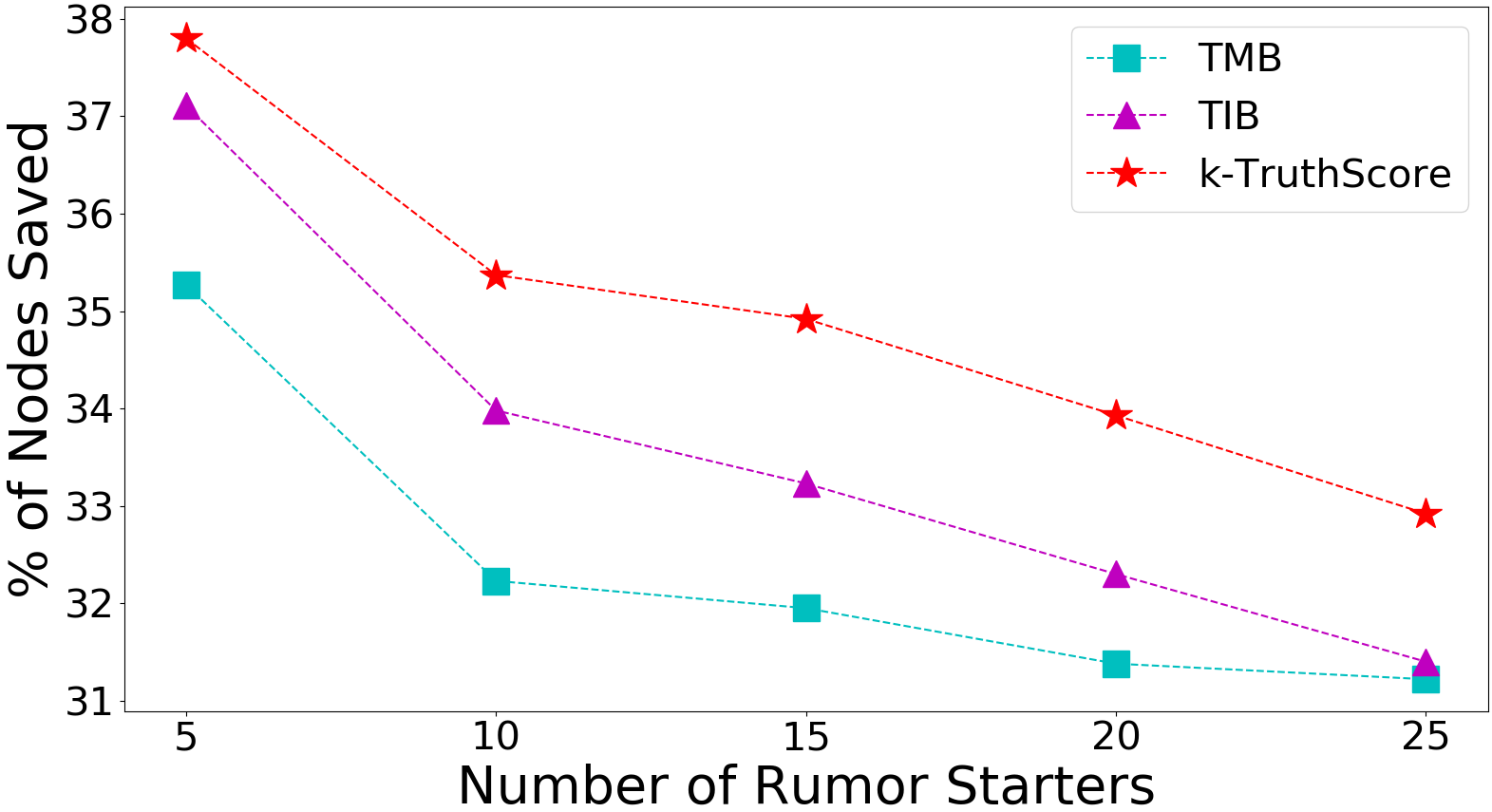}
	\begin{center} { (b) Facebook}\end{center}		
	\includegraphics[width=\linewidth]{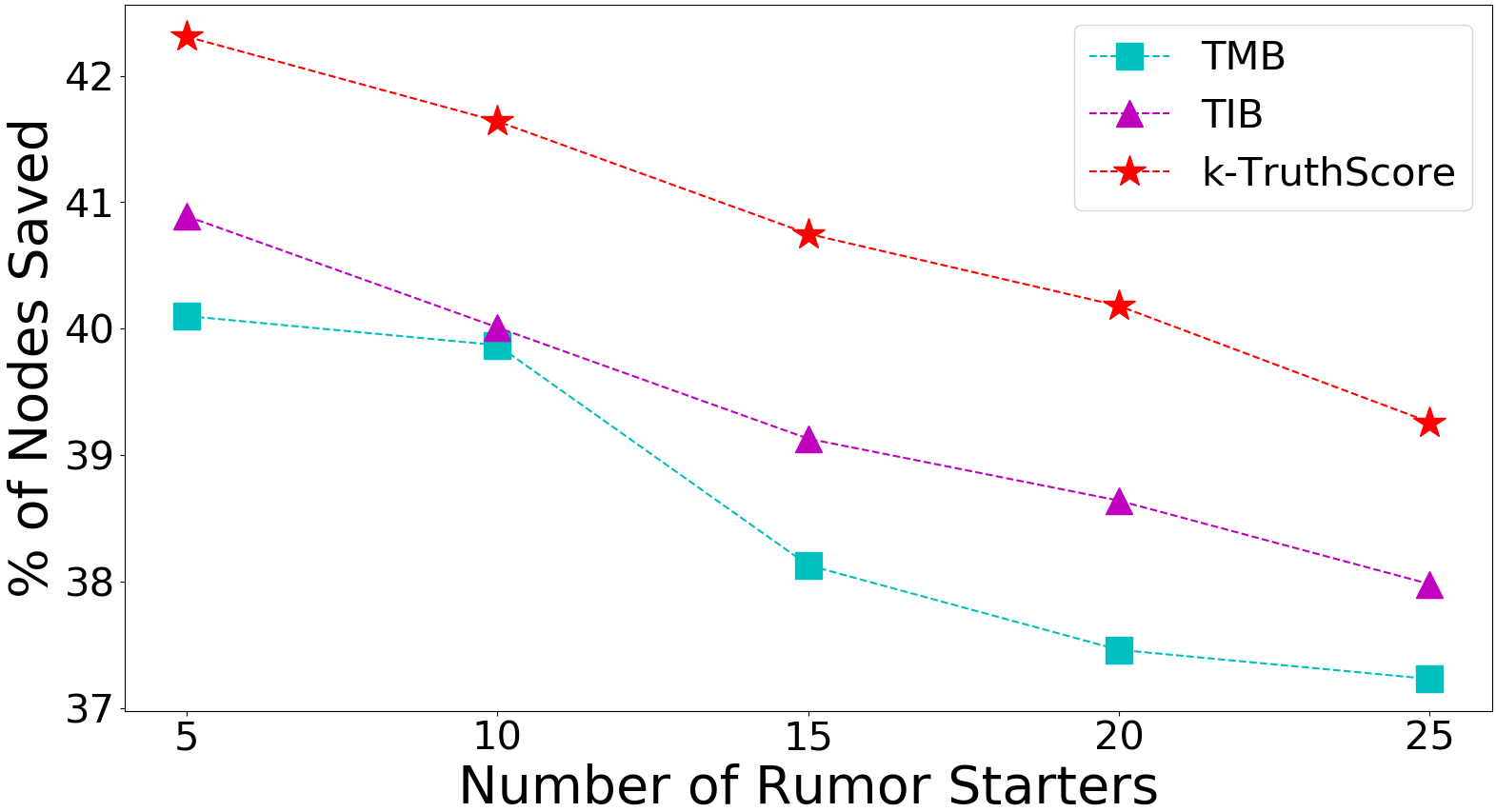}
	\begin{center} { (c) Twitter}\end{center}
\caption{Effect of varying $|M|$ when $k$=5.}\label{fig1b}
\end{minipage}
\end{figure*}

First, we study the performance of $k-TruthScore$ as a function of chosen truth-campaigners $k$, varying $k$ from $2$ to $10$. We also set the deadline for the misinformation to be the network diameter, if not specified otherwise. Figure~\ref{fig1a} shows that the $k-TruthScore$ outperforms state-of-the-art methods for finding the top-$k$ truth-campaigners. TIB and TMB methods are designed to choose truth-campaigners globally, and we restrict these methods to choose truth-campaigners from the given set of prospective truth-campaigners. Under this restriction, $k-TruthScore$ significantly outperforms both TIB and TMB methods.

\begin{figure*}
\centering
\begin{minipage}[t]{.47\textwidth}
\centering
	\includegraphics[width=\linewidth]{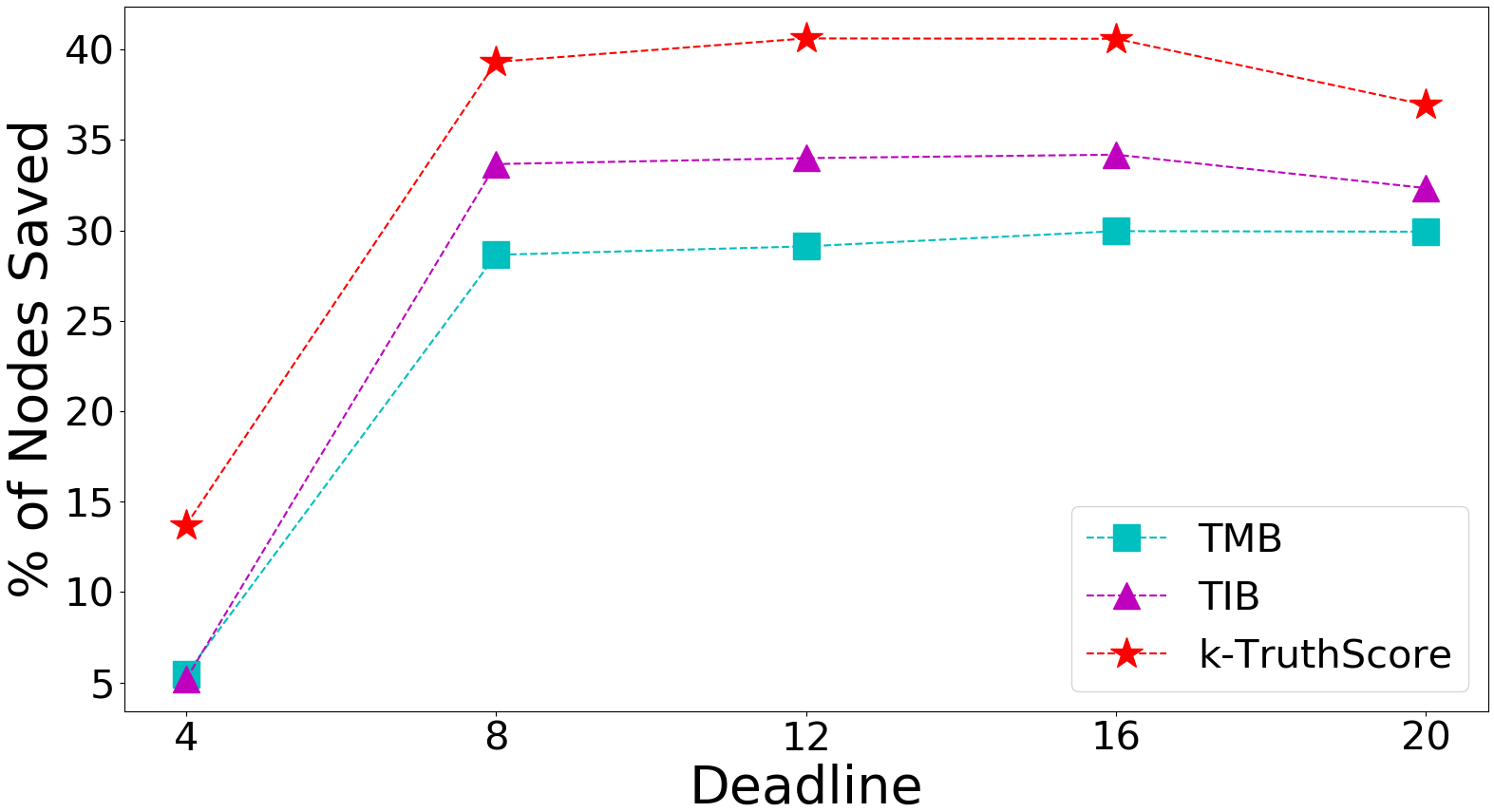}
	\caption{Effect of varying deadline $\alpha$ on Digg Network.}
	\label{fig2a}
\end{minipage}\qquad
\begin{minipage}[t]{.47\textwidth}
\centering
	\includegraphics[width=\linewidth]{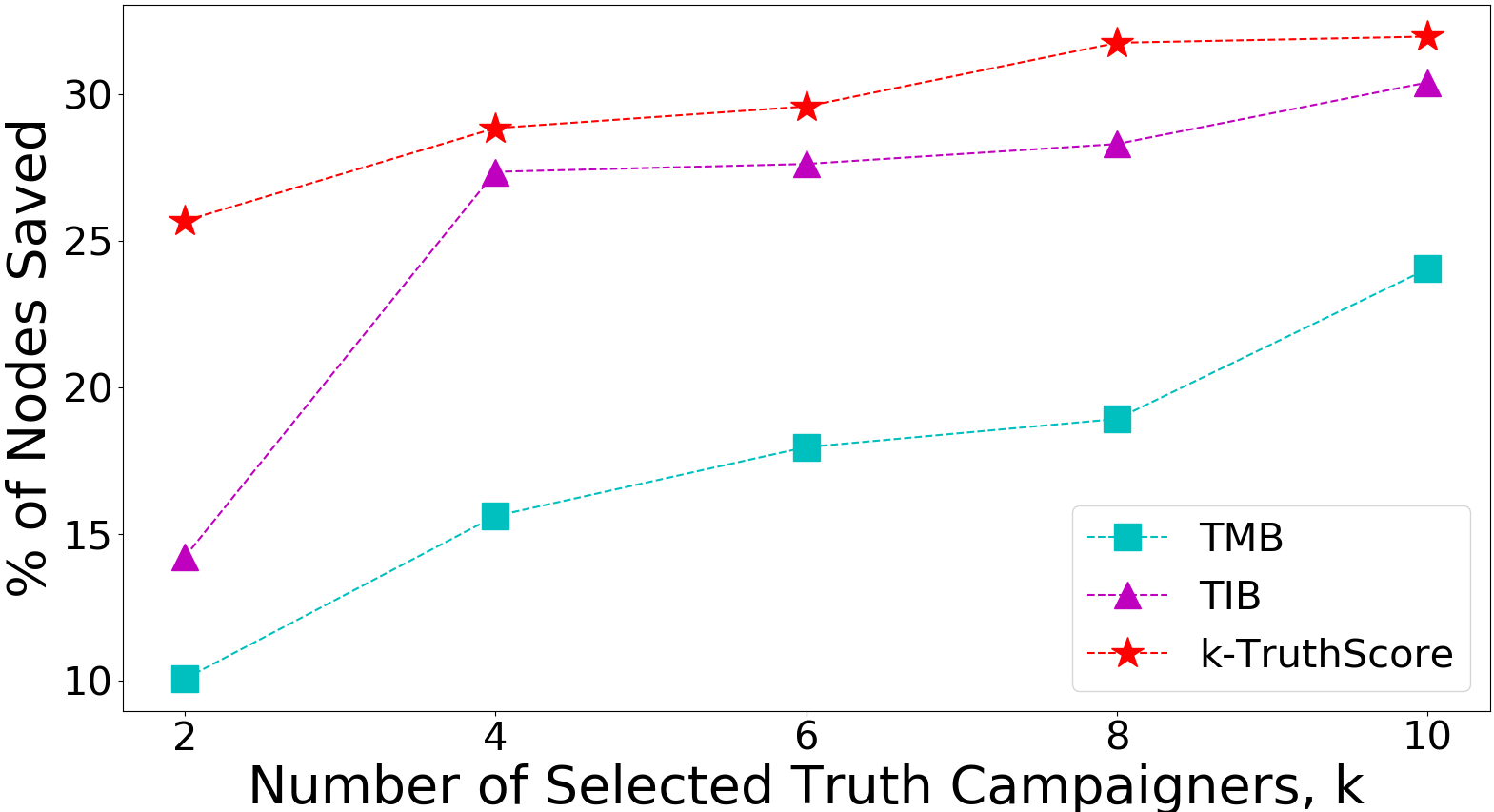}
	\caption{Square-bias function on Digg Network.}
	\label{fig2b}
\end{minipage}
\end{figure*}

Next, we study the impact of varying the number of rumor starters. We fix $k = 5$, and allow $M$ to vary from 5 to 25. Figure~\ref{fig1b} shows that in this case, the percentage of nodes saved reduces as the number of rumor starters increases, while $k-TruthScore$ still outperforms TIB and TMB methods.

We also study the impact of varying the deadline on the percentage of nodes saved when $|M|=10$ and $k=5$. The results are shown for the Digg network and $\alpha$ varying from 4 to 20. The results show that the percentage of the saved nodes is maximum around the iteration time $\alpha = diameter(G)$, and consistently $k-TruthScore$ outperforms at any other set deadline. 

The results discussed so far depend on linear degradation, as presented in Section~\ref{sec:cicmb}. We now study the efficiency of $k-TruthScore$ for a quadratic bias function. We evaluate $k-TruthScore$ for the following bias function:\\
When $u$ changes its state to $M$, $B_m(u)[i+1]=B_m(u)[i]$ $\&$ $B_t(u)[i+1]=({B_t(u)}[i])^2$.\\
When $u$ changes its state to $T$, $B_m(u)[i+1]=({B_m(u)}[i])^2$ $\&$ $B_t(u)[i+1]=B_t(u)[i]$.

The results show that $k-TruthScore$ saves the maximum number of nodes for the quadratic bias function. However, the percentage of saved nodes is smaller than the ones observed in Figure 3. This is due to the reason that the square function reduces the biases faster, and users are more stubborn to change their state once they have believed in one information.

\section{Conclusion} \label{conclusion}
The current research presents a solution to the problem of minimizing the impact of misinformation by propagating its counter true information in OSNs. In particular, we look to identify the top-$k$ candidates as truth-campaigners from a given set of prospective truth-campaigners and given rumor starters. We first propose a propagation model called the Competitive Independent Cascade Model with users' Bias that considers the presence of strong user bias towards different opinions, believes, or political parties. For our experiments, we used two different functions to capture the bias dynamics towards true and misinformation, one linear and one quadratic.

Next, we introduce an algorithm, $k-TruthScore$, to identify top-$k$ truth-campaigners, and compare the results against two state of the art algorithms, namely Temporal Influence Blocking and Targeted Misinformation Blocking. To compare the algorithms, we compute the percentage of saved nodes per network $G$, by the deadline $\alpha$. A node is tagged as saved if at the deadline $\alpha$ it believes in true information and would have believed in the misinformation in the absence of truth campaigners. 

We compare the three algorithms under each of the two bias functions on three different networks, namely Digg, Facebook, and Twitter.  Moreover, we compare the three algorithms by varying the number of originating rumor spreaders as well as varying the deadline at which we compute the TruthSocre. Our results show that $k-TruthScore$ outperforms the state of the art methods in every case. 

In the future, we would like to do an in-depth analysis of the CICMB model for different bias functions, such as constant increase/decrease (where the bias values are increased or decreased by a constant value, respectively), other linear functions (for example, if one bias value of a user increases then the other decreases), different quadratic functions, and so on. The proposed $k-TruthScore$ method outperforms for both the considered functions; however, one can propose a method, i.e., specific to a given bias function.

\bibliographystyle{splncs04}
\bibliography{mybib.bib}

\end{document}